\documentclass[11pt]{article}
\usepackage[utf8]{inputenc}
\usepackage[T1]{fontenc}
\usepackage{lmodern}
\usepackage[margin=1in]{geometry}
\usepackage{amsmath,amssymb}
\usepackage{graphicx}
\usepackage{booktabs}
\usepackage{hyperref}
\usepackage{xcolor}
\usepackage{natbib}
\usepackage{enumitem}
\usepackage{placeins}
\usepackage{url}
\usepackage{microtype}
\hypersetup{colorlinks=true, linkcolor=blue!70!black, citecolor=blue!70!black, urlcolor=blue!70!black}
\newcommand{\sstar}{s^\star}
\newcommand{\Nexp}{N_{\mathrm{exp}}}

\title{Caching for Dollars, Not Hits:\\ An Exact Offline Reference for Cloud-Egress Caching\\ and the Crossover That Decides When It Pays}
\author{Madhulatha Mandarapu\thanks{madhulatha@samyama.ai} \and Sandeep Kunkunuru\thanks{sandeep@samyama.ai}}
\date{VaidhyaMegha Private Limited, India\\[2pt]\url{https://samyama.ai/}\\[8pt]June 2026}

\begin{document}
\maketitle

\begin{abstract}
When a cache miss fetches from cloud object storage, the bill is per GET request and per byte of egress,
not latency. Classic caching minimizes the miss rate, the wrong objective: a rarely but expensively
fetched object can cost thousands of times more dollars than a frequently but cheaply fetched one.
Generalized-caching theory bounds the miss-cost objective, but no reported benchmark measures how far
deployed heuristics sit from the dollar-optimal offline policy on real cloud prices. We supply that
reference. For uniform-size page caches with heterogeneous miss costs the offline dollar-optimum is exact
in polynomial time via an integral interval linear program -- validated against brute force; variable
sizes are NP-hard, so we extend the flow-based offline bound from the hit-ratio objective to
dollars (cost-FOO), tight to about four percent. Against this reference we find: (i) a heterogeneity-regret
law -- LRU's dollar-regret rises with miss-cost dispersion (Spearman 0.87) while cost-aware GreedyDual cuts
it to roughly a tenth; (ii) a contention frontier -- GreedyDual's residual regret collapses to near zero
exactly when the budget fits the expensive working set, and is the open slice otherwise; and (iii) a
closed-form crossover $\sstar = \text{GET\_fee}/\text{egress\_rate}$ (about 4\,KB on S3, 330\,B on GCS) that
predicts which deployments need dollar-aware caching. On real memcache and CDN traces the price vector alone
moves the workload across $\sstar$, shifting the regime as predicted. The artifact is a reproducible
billing-faithful benchmark; heuristics and bounds it builds on are prior work, credited.
\end{abstract}

\section{Introduction}
A compute node caches objects locally; a miss fetches from cloud object storage (S3, GCS, Azure Blob),
which bills \emph{per GET request} and \emph{per byte of egress / cross-zone transfer}, not in latency.
The miss cost of object $i$ is therefore
\begin{equation}\label{eq:cost}
  c_i = \underbrace{f}_{\text{GET fee}} + \underbrace{s_i\,e}_{\text{egress}} \;(+\,\text{latency penalty}),
\end{equation}
with size $s_i$ and per-byte egress rate $e$. This cost is \emph{heterogeneous}: a large or cross-region
object costs orders of magnitude more than a small same-region one, and the flat GET fee matters
independently of bytes. Minimizing the miss \emph{rate} -- the classic objective behind LRU and
Belady's rule~\citep{belady1966} -- is then the wrong objective. A one-slot cache with a 1\,KB object
accessed 100 times and a 1\,GB object accessed 10 times illustrates it: hit-rate caching keeps the small
hot object and saves $\sim\,\$5\times10^{-5}$; keeping the large cold object saves $\sim\,\$0.90$, over
four orders of magnitude more dollars for fewer hits.

Minimizing total billed dollars under a capacity budget is \emph{generalized (size- and cost-weighted)
caching}. Its online competitive theory is settled up to constants: $O(\log k)$ randomized and $k$
deterministic~\citep{bansal2012generalized}, with the GreedyDual-Size family the practical
workhorse~\citep{cao1997greedydual} and learning-augmented variants for predicted
reuse~\citep{lykouris2021predictions,antoniadis2020untrusted}. What is missing is empirical: \emph{no
reported benchmark measures how far deployed heuristics sit from the dollar-optimal offline policy on real
cloud price structures}. That gap is a measurement gap, and we close it.

\paragraph{Contributions.}
\begin{enumerate}[leftmargin=1.5em]\itemsep1pt
  \item \textbf{An exact dollar-optimal reference.} For uniform-size page caches with heterogeneous costs,
        the offline dollar-optimum is an \emph{integral} interval LP, hence exact in polynomial time
        (\S\ref{sec:model}); validated to the cent against brute force. Variable sizes are
        NP-hard~\citep{folwarczny2015general}, so we extend the flow-based offline bound (FOO)
        \citep{berger2018foo} from hit-ratio to dollars (cost-FOO), tight to $\approx 4\%$.
  \item \textbf{A heterogeneity-regret law and a contention frontier} (\S\ref{sec:exp}): cost-blindness
        (LRU) costs more as miss-cost dispersion $H$ grows; cost-awareness converts that into a
        \emph{contention} problem whose residual regret collapses exactly when the budget fits the
        expensive working set ($B=\Nexp$).
  \item \textbf{The crossover $\sstar=f/e$} (\S\ref{sec:cross}): a closed-form, price-vector rule for
        \emph{when} dollar-aware caching pays, validated on a real production trace.
\end{enumerate}
We add no new caching algorithm and beat no competitive ratio; the contribution is the exact reference and
the characterization it enables. Code, data scripts, and the full pre-registration are
public.\footnote{\url{https://github.com/samyama-ai/cloud-egress-cache}}

\section{Model and the exact dollar-optimum}\label{sec:model}
A cache of capacity $B$ serves a request stream over objects with sizes $s_i$ and miss costs $c_i$
\eqref{eq:cost}; total cost is the sum of $c_i$ over every fetch. We score policies in \emph{dollars}.

\paragraph{Interval-packing optimum.} For each request $t$ whose object recurs at $\mathrm{next}(t)$, a
binary $x_t$ decides whether the object is retained across the gap, yielding a hit (saving $c_{o(t)}$) and
occupying a slot at every interior step. The dollar-optimum maximizes savings subject to, at each step
$\tau$,
\begin{equation}\label{eq:lp}
  s_{o(\tau)} + \!\!\sum_{t:\,t<\tau<\mathrm{next}(t)}\!\! s_{o(t)}\,x_t \;\le\; B .
\end{equation}
Each variable covers a \emph{contiguous} range of constraints, so for \textbf{uniform sizes} the
constraint matrix has the consecutive-ones property and is totally unimodular: the LP relaxation is
integral and \eqref{eq:lp} solves the offline dollar-optimum \emph{exactly} in polynomial time. We verify
this is cent-exact against an exhaustive brute force on $250$ random instances. Because the constraints are
intervals, the same optimum is a \emph{min-cost flow} on the time line (a $B{-}1$-capacity ``shelf'' path
with one unit-capacity arc per gap, cost $-c_i$); this equivalent form -- validated to match the LP and
brute force -- scales the exact optimum past the dense LP to $10^5$ requests, which we use to check that the
real-trace regret is scale-stable. For \textbf{variable
sizes}, the size weights break unimodularity; the LP relaxation is a fractional-caching \emph{lower bound}
on the NP-hard optimum~\citep{folwarczny2015general} -- the dollar analogue of FOO~\citep{berger2018foo},
which bounds the hit-ratio optimum. A feasible policy upper-brackets it; we call the pair \emph{cost-FOO}.
This exact-or-tightly-bounded reference is what lets us measure \emph{regret}, $R(\pi)=(\mathrm{Cost}(\pi)-\mathrm{Cost}(\mathrm{OPT}))/\mathrm{Cost}(\mathrm{OPT})$, against a true optimum rather than against another heuristic.

\paragraph{Policies.} We score LRU, LFU, GreedyDual-Size with cost (GDS) and its frequency variant
(GDSF)~\citep{cao1997greedydual}, Belady (hit-rate oracle)~\citep{belady1966}, and a cost-aware Belady
heuristic -- all in dollars.

\section{The GET-fee / egress crossover}\label{sec:cross}
Equation~\eqref{eq:cost} has a scale at which the flat fee and the egress term are equal:
\begin{equation}\label{eq:cross}
  \sstar \;=\; f/e .
\end{equation}
Below $\sstar$ a miss is dominated by the GET fee, so costs are near-homogeneous and hit-rate caching is
near-optimal; above $\sstar$ egress dominates, costs are heterogeneous, and cost-aware caching matters.
This is a property of the \emph{price vector}, not the workload: with list prices (June 2026) $\sstar\approx
4.4$\,KB on S3 internet egress, $\approx 330$\,B on GCS, $\approx 460$\,B on Azure, and $\approx 20$\,KB on
S3 cross-region transfer. It predicts \emph{which deployments need dollar-aware caching at all}.

\section{Experiments}\label{sec:exp}
All hypotheses and thresholds were pre-registered before the runs; one pre-registered regime form failed
and was reframed (below), disclosed in the public pre-registration. Synthetic workloads use Zipf
popularity assigned independently of size, so cheap-hot versus expensive-cold tension exists.

\begin{figure}[t]\centering
  \begin{minipage}{0.48\linewidth}\includegraphics[width=\linewidth]{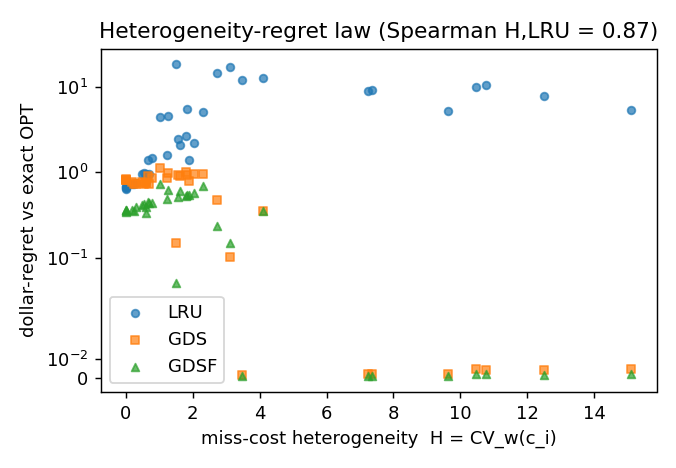}
    \caption{Heterogeneity-regret law: dollar-regret versus miss-cost heterogeneity $H$ (exact OPT).}
    \label{fig:law}\end{minipage}\hfill
  \begin{minipage}{0.48\linewidth}\includegraphics[width=\linewidth]{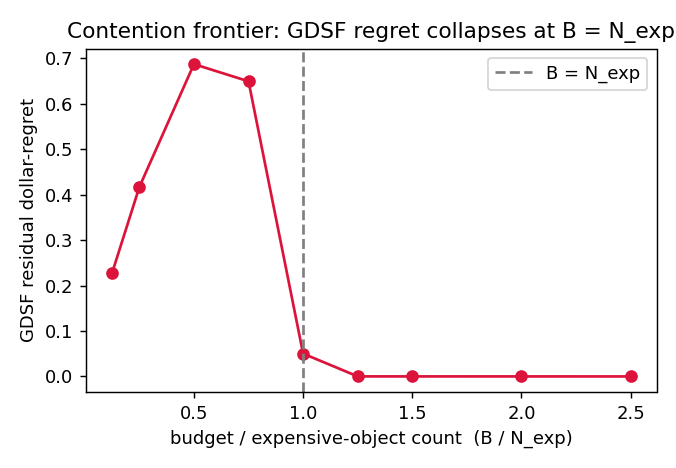}
    \caption{Contention frontier: GDSF residual regret collapses at $B=\Nexp$.}\label{fig:cont}\end{minipage}
\end{figure}

\paragraph{Heterogeneity-regret law (Fig.~\ref{fig:law}).} Let $H$ be the access-weighted coefficient of
variation of the miss-cost vector. LRU's dollar-regret rises with $H$ (Spearman $0.87$); cost-aware GDSF's
median regret is $0.13\times$ LRU's where $H\ge 0.5$. Cost-blindness is expensive, and cost-awareness
buys most of it back.

\paragraph{Contention frontier (Fig.~\ref{fig:cont}).} The pre-registered ``three regimes in $H$'' form
\emph{failed}: even at $H=0$, LRU carries an intrinsic recency regret ($\sim$0.65 vs Belady), so raw LRU
regret conflates recency and egress components. The honest structure is two knobs. Heterogeneity drives
LRU's cost-blindness, which GDSF removes; the \emph{residual} is governed by expensive-object contention.
With $\Nexp$ expensive objects, GDSF's regret is $0.23$--$0.69$ while $B<\Nexp$ and collapses to $0.0002$
\emph{exactly} at $B=\Nexp$: once the expensive working set fits, greedy cost-ranking is optimal; when it
does not, greedy provably leaves money on the table -- the open slice. Cost-awareness converts a
heterogeneity problem into a contention problem.

\paragraph{cost-FOO bracket.} On variable-size synthetic traces the cost-FOO bracket $(U-L)/L$ is a median
$0.04$, so variable-size regret numbers are meaningful rather than artifacts of a loose bound.

\begin{figure}[t]\centering
  \begin{minipage}{0.48\linewidth}\includegraphics[width=\linewidth]{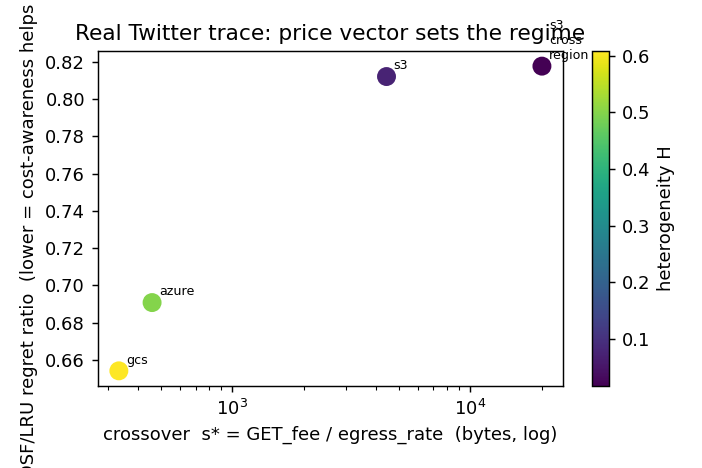}
    \caption{Real Twitter memcache trace ($H<1$): the price vector, through $\sstar$, sets the
      regime.}\label{fig:real}\end{minipage}\hfill
  \begin{minipage}{0.48\linewidth}\includegraphics[width=\linewidth]{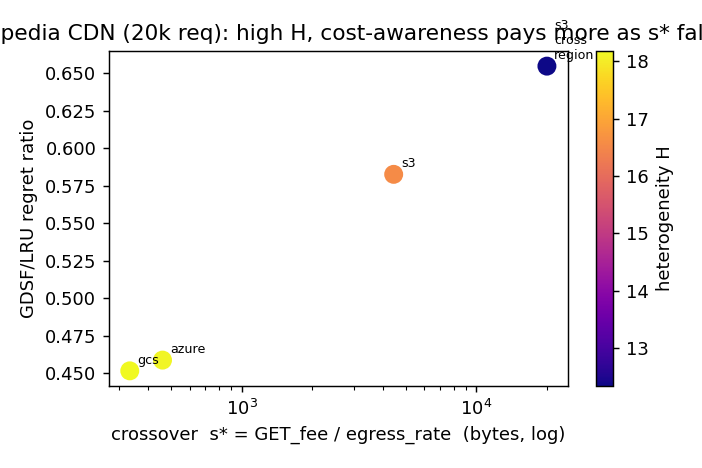}
    \caption{Real Wikipedia CDN trace ($H=12$--$18$): GDSF/LRU regret ratio falls as $\sstar$
      drops.}\label{fig:cdn}\end{minipage}
\end{figure}

\paragraph{Real small-object trace (Twitter memcache; Table~\ref{tab:real}, Fig.~\ref{fig:real}).} We replay
a real Twitter twemcache production trace~\citep{yang2020twitter} (cluster 52; a $20{,}000$-request window,
real per-object sizes), computing the exact dollar-optimum, under four real price vectors. Under S3 the small
memcache objects (mean $243$\,B) sit \emph{below} $\sstar\approx 4.4$\,KB, so misses are GET-fee-dominated:
$H=0.075$ and GDSF $\approx$ LRU -- egress-aware caching buys almost nothing on small-object cache traffic,
a useful negative. The \emph{same} trace under GCS or Azure pricing (a $10\times$ cheaper GET fee pushes
$\sstar$ to $\sim$330--460\,B) crosses the threshold: $H$ rises to $0.5$--$0.6$ and the GDSF/LRU regret
ratio falls from $0.82$ to $0.65$. The regime is set by the price vector, exactly as $\sstar$ predicts.

\begin{table}[t]\centering\small
\begin{tabular}{lrrrr}
\toprule
price vector & $\sstar$ (bytes) & $H$ & LRU regret & GDSF/LRU \\
\midrule
S3 cross-region & 20000 & 0.017 & 0.152 & 0.82 \\
S3 internet     & 4444  & 0.075 & 0.153 & 0.81 \\
Azure internet  & 460   & 0.498 & 0.160 & 0.69 \\
GCS internet    & 333   & 0.608 & 0.162 & 0.65 \\
\bottomrule
\end{tabular}
\caption{Same real Twitter trace, four real price vectors. As the crossover $\sstar$ falls, more objects
become egress-dominated, $H$ rises, and cost-aware caching helps more.}\label{tab:real}
\end{table}

\paragraph{Real large-object trace (Wikipedia CDN; Fig.~\ref{fig:cdn}).} We replay the Wikipedia CDN
trace~\citep{song2020lrb} (mean object $37$\,KB, max $94$\,MB -- half the objects exceed $\sstar$), so
$H=12$--$18$, deep in the heterogeneous regime. The exact dollar-optimum is computed by the min-cost-flow
solver. The actionable signal of the small-object arm \emph{repeats and strengthens}: as $\sstar$ falls
across the four price vectors, the GDSF/LRU regret ratio drops monotonically $0.65\to0.45$ -- cost-awareness
captures more of the recoverable dollar-regret as egress dominates. Two honest caveats. First, absolute LRU
regret is \emph{modest} ($3$--$7\%$): CDN traffic has low reuse (a long tail of one-hit-wonders that no
policy can cache), and the largest, most expensive objects are disproportionately such single-touch fetches,
so much billed cost is unavoidable for every policy. Second, computing the exact optimum at $5\times$ the
window (\,$10^5$ requests) leaves LRU's regret unchanged, so the windowed number is representative, not an
artifact. The strong ``leaves money on the table'' regime (\S contention) needs heterogeneity \emph{and}
high reuse \emph{and} a tight budget simultaneously; neither real trace we tested hits all three -- itself a
useful finding about where dollar-aware caching does and does not move the bill.


\FloatBarrier
\section{Related work}\label{sec:related}
Cost-aware caching by miss-cost-over-size weighting is GreedyDual-Size~\citep{cao1997greedydual}, the
practical workhorse we measure. The online competitive theory for generalized (size- and cost-) caching is
$O(\log k)$ randomized~\citep{bansal2012generalized}; ski-rental / rent-or-buy~\citep{karlin1994skirental}
governs the per-request-fee batching sub-problem; learning-augmented caching adds predicted
reuse~\citep{lykouris2021predictions,antoniadis2020untrusted}. The offline side is where our reference
lives: general caching with variable sizes is NP-hard~\citep{folwarczny2015general}, and FOO/PFOO
\citep{berger2018foo} compute tight flow-based bounds on the \emph{hit-ratio} optimum -- we extend the
construction to the \emph{billing} objective. Cloud engines cache object-store results
aggressively~\citep{dageville2016snowflake}, but evaluations are workload-specific with no dollar-optimality
reference. Our real arms use the Twitter production cache traces~\citep{yang2020twitter} and the Wikipedia
CDN trace~\citep{song2020lrb}; the classic hit-rate oracle is Belady's rule~\citep{belady1966}.

\section{Limitations and honest scope}\label{sec:limits}
This is a reproducible benchmark and characterization, not a new caching algorithm. The pre-registered
three-regime-in-$H$ form failed and was reframed into the two-knob decomposition above; we report the
failure rather than relabel it. The dense interval-LP optimum does not scale (millions of nonzeros at tens
of thousands of requests); the min-cost-flow form pushes the \emph{exact} optimum to $10^5$ requests, enough
to confirm scale-stability, but full production scale ($10^7$+) needs a faster flow solver -- the remaining
scalability gap. Both real traces we tested leave LRU's dollar-regret modest (memcache: GET-fee-dominated;
CDN: low reuse), so the large ``leaves-money'' regret is so far a synthetic-regime result; finding a real
workload with high heterogeneity \emph{and} high reuse \emph{and} budget pressure is open. List prices are
date-stamped and re-tiering shifts $\sstar$; adversarial re-pricing is left to future work. Every number is
regenerated by one command over public code.

\small
\bibliographystyle{plainnat}
\bibliography{paper13_egress_caching}
\end{document}